\documentclass[aps,prl,twocolumn,superscriptaddress,showpacs]{revtex4-1}

\usepackage{graphicx}
\usepackage{amssymb}
\usepackage{color}

\begin{document}


\title{Quantum Critical Spin-2 Chain with Emergent SU(3) Symmetry}

\author{Pochung Chen}
\email[]{pcchen@phys.nthu.edu.tw}
\affiliation{Department of Physics and Frontier Research Center on Fundamental and Applied Sciences of Matters, National Tsing Hua University, Hsinchu 30013, Taiwan}
\affiliation{Physics Division, National Center for Theoretical Sciences, Hsinchu 30013, Taiwan}

\author{Zhi-Long Xue}
\affiliation{Department of Physics and Frontier Research Center on Fundamental and Applied Sciences of Matters, National Tsing Hua University, Hsinchu 30013, Taiwan}

\author{I. P. McCulloch}
\affiliation{Centre for Engineered Quantum Systems, The University of Queensland, Brisbane, Queensland 4072, Australia}

\author{Ming-Chiang Chung}
\affiliation{Physics Division, National Center for Theoretical Sciences, Hsinchu 30013, Taiwan}
\affiliation{Department of Physics, National Chung Hsing University, Taichung 40227, Taiwan}ß

\author{Chao-Chun Huang}
\author{S.-K. Yip}
\affiliation{Institute of Physics and Institute of Atomic and Molecular Sciences,
 Academia Sinica, Taipei 11529, Taiwan}


\date{\today}

\begin{abstract}
We study the quantum critical phase of an SU(2) symmetric spin-2 chain obtained from spin-2 bosons  in a one-dimensional lattice.
We obtain the scaling of the finite-size energies and entanglement entropy by exact diagonalization
and density-matrix renormalization group methods.
From the numerical results of the energy spectra, central charge, and scaling dimension
we identify the conformal field theory describing the whole critical phase to be the SU(3)$_1$ Wess-Zumino-Witten model.
We find that, while  the Hamiltonian is only SU(2) invariant, in this critical phase
there is an emergent SU(3) symmetry in the thermodynamic limit.
\end{abstract}


\maketitle



Cold atomic gases in optical lattices have become an ideal framework for studying
quantum many-body systems in recent years \cite{Bloch:2008gl}.
In particular, various schemes have been proposed to study quantum magnetism \cite{Lewenstein:2007hr}.
For spin 1/2 systems, simulation of the Ising model has been realized using boson
in a tilted optical lattice \cite{Simon:2011hu}. It has also been proposed that the spin 1/2 XYZ Heisenberg model
can be realized using $p$-orbital bosons \cite{Pinheiro:2013cd}. This rapid progress in cold atomic physics
results in a considerable renewal of interest to study models with higher spins or higher symmetries,
especially for models which are potentially realizable by cold atomic systems.
A natural direction is to study spinor bosons and their novel phases.
For example, it has been proposed that the spin-1 bi-linear bi-quadractic (BB) model can be engineered
using spin-1 cold bosons in optical lattice \cite{Yip:2003dj,Imambekov:2003jr}.
Furthermore, the phase diagram of spin-1 bosons in one-dimensional (1D) lattice
has been studied numerically and compared to the spin-1 BB model \cite{Rizzi:2005gw}.
Since the spin-2 bosons are available and have been experimentally studied \cite{Schmaljohann:2004bl,Kuwamoto:2004bk,Widera:2006cr,Tojo:2009hb},
it is of great interest to explore the phases realizable by spin-2 bosons.
On the other hand, it has also been pointed out that fermions with hyperfine spin $F=3/2$ can be used
to realize models with SO(5) symmetry \cite{CONGJUNWU:2011gm},
or to realize SU(3) spin chain by effectively suppressing
the occupation of one of the middle states \cite{Greiter:2007bm}.
Possibilities to realize higher SU(N)  symmetry have also been proposed \cite{Cazalilla:2009wl,Messio:2012bc}.
Along this line, spin dynamics and correlation have been studied experimentally
using cold fermions with effective spin ranging from 1/2 to 9/2 \cite{Krauser:2012iv,Greif:2013kb}.
Another interesting question is to explore symmetries that emerge in the low energy limit of the models.
Indeed, different aspects of emergent symmetries have been discussed widely in the recent literature.  
Examples include  SO(5) and SO(8) symmetries in high temperature superconductors and two-leg ladders \cite{SO5,SO8},
$E_8$ symmetry in Ising spin chains under a critical transverse field \cite{E8},
emergent modular and translational symmetries for quantum Hall states \cite{QHE} and fractional Chern insulator \cite{FCI},
supersymmetry at sample boundaries of topological phases \cite{super}, 
or at critical or multicritical points separating different phases \cite{SO8},
especially for confinement-deconfinement or non-Landau phase transitions \cite{deconf}.

Recently we studied the phase diagram of spin-2 bosons in a 1D optical lattice
with one particle per site and identify three
possible phases for a finite system: ferromagnetic, dimerized, and trimerized phases \cite{Chen:2012ft}.
Within the trimerized phase, if the system size is a multiple of three,
the ground state is a spin-singlet with finite-size gap and broken lattice symmetry.
It was also shown that in the thermodynamic limit the system became gapless with unbroken lattice symmetry.
The nature of this extended critical phase was, however, not fully determined.
In this letter we investigate the extended critical phase of the spin-2 boson in a 1D lattice.
In particular, we identify the conformal field theory (CFT) describing the low energy physics
of the whole critical phase.
By using multiple diagnostic tools we show that in the thermodynamic limit the low energy physics of
this critical phase is described by the SU(3)$_1$ Wess-Zumino-Witten (WZW) model. This is the main result of this work.

We begin with the Hamiltonian, which is obtained from spin-2 bosons in a 1D optical lattice
with one particle per site in the limit of $t/U_S \ll 1$.
Here $t$ is the hopping between nearest neighbor and $U_S$ is the Hubbard repulsion
for two particles with spin $S$ on the same site.  Within second order perturbation theory
the effective Hamiltonian reads
\begin{equation}
  H=\sum_i H_{i,i+1}=\sum_i \epsilon_o P_{0,i,i+1}+\epsilon_2 P_{2,i,i+1}+\epsilon_4 P_{4,i,i+1},
\end{equation}
where $\epsilon_S = - 4 t^2/U_S$. Here $i$ is the site index,
$P_{S,i,i+1}$  denotes the projection operator for sites $i$ and $i+1$ onto a state with total  spin $S$.
We focus on the regime with $U_S>0$ (hence $\epsilon_S < 0$) to ensure
stability for one-particle per site.
We use non-Abelian density matrix renormalization group (DMRG) that preserves the SU(2) symmetry.
This not only increases the accuracy and but also allows us to target any specific total SU(2) spin sector.
Within DMRG, however, it is more convenient to explicitly express $H_{i,i+1}$
in terms of spin-2 operators $\mathbf{S}_i$, resulting
$ H_{i,i+1}=\sum_{n=1}^4   \alpha_n(\epsilon_0, \epsilon_2, \epsilon_4) \left(\mathbf{S}_i \cdot \mathbf{S}_{i+1} \right)^n$,
where the expressions of $\alpha_n$ can be found in Ref~\cite{Chen:2012ft}.
While the mean-field and exact phase diagrams have been studied \cite{Ciobanu:2000jz,Chen:2012ft},
the nature of the critical phase is not known.
In the following we will use the finite-size scaling of the energies and the entanglement entropy to identify the CFT.
Since the whole critical phase is described by a unique CFT,
it suffices to use one particular parameter set for the determination.
Throughout this paper, we will use  $\vec{\epsilon}\equiv(\epsilon_0,\epsilon_2,\epsilon_4)=(0,-1,0)$.
This sets the system deep in the trimerized phase and far from the boundary of the ferromagnetic and dimerized phases.

\begin{figure}
  \includegraphics[width=0.9\columnwidth]{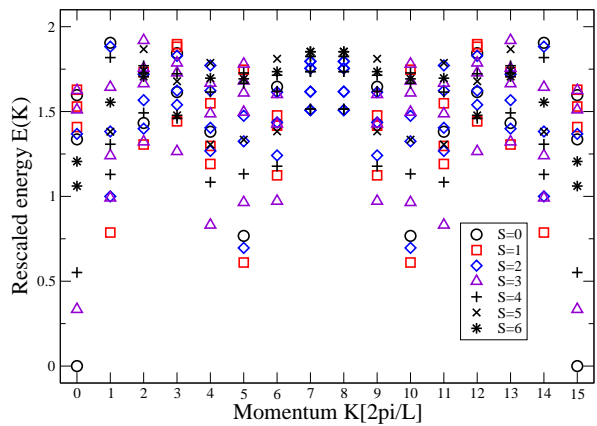}
  \caption{\label{fig:ED_L15}
  (Color online) Rescaled energy spectrum obtained from exact diagonalization with $L=15$ and $\vec{\epsilon}=(0,-1,0)$.
  }
\end{figure}


We start with the exact diagonalization (ED) to obtain the low energy spectrum of small size systems with periodic boundary conditions.
In Fig.~\ref{fig:ED_L15} we show the excitation spectrum for $L=15$.
We set the ground state energy to be zero and use the energy of the $S=2$ state at $k=2\pi/L$
as the energy unit (for reason described later in the text).
We find that the ground state has $S=0$ when the system size is a multiple of three.
Furthermore, we observe that there is a period-three structure and soft-modes develop at
 $k=\pm 2\pi/3 ({\rm mod} 2 \pi)$,
with a cluster of low energy states with $S=0,1,2$.
The period-three structure and the fact that gapped trimerized state is formed for finite-size chain
strongly suggests that the critical theory has approximate SU(3) symmetry at finite sizes.
This observation above allows us to rule out some models as the potential CFT of the critical phase.
Since our Hamiltonian is SU(2) symmetric, the corresponding CFT must contain SU(2) as a subgroup.
SU$_k$(2) models are natural candidates, however, their spectrum would have minima at $k = \pi$ rather than
$\pm 2 \pi/3 $, and the spin correlations have period-two rather than period-three.
Consequently, one can rule out SU$_k$(2) models as the associated CFT.
On the other hand, the low energy spectrum is compatible with DMRG calculation results
for the SU(3) Heisenberg model in Ref.~\cite{Aguado:2009dma}
and ED results of spin-1 BB model in the critical period-three phase \cite{sup}.
This makes SU(3)$_1$ WZW model with central charge $c=2$ an appealing candidate.

For SU(3)$_1$ WZW model the soft modes at $k=\pm 2\pi/3$ should have degeneracy that
matches the dimension of SU(3) representation. Here because the bare Hamiltonian has SU(2)
but not the SU(3) symmetry, for finite-size system it is natural that the energies at $k=\pm 2\pi/3$ will split according to
the SU(2) spin as observed in Fig.~\ref{fig:ED_L15}.
However, the $S=3, 4$ states at $k=0$ have lower energies than the  states at $k=\pm 2\pi/3$.
These low energy states are not expected if the critical theory is SU(3)$_1$ WZW model.
We shall argue that the presence of these states are due to the proximity to the ferromagnetic
phase and the lack of SU(3) symmetry in the Hamiltonian. We shall provide more details on this point below.
Since the SU(3) is only emergent, it is then not surprising that non-CFT behavior is observed for small sizes.
However we expect that for sufficiently large sizes, their re-scaled energies will move up
and the low energy spectrum will become fully consistent with the CFT.
In the following we shall use DMRG to calculate the energies of the states that are consistent with the
SU(3)$_1$ WZW model. The finite-size scaling of these energies then are used to identify the CFT.
While we can no-longer specify the momenta of the excited states in DMRG,
we can target particular SU(2) spin sector to obtain the low energy states needed.

\begin{figure}
  \includegraphics[width=0.9\columnwidth]{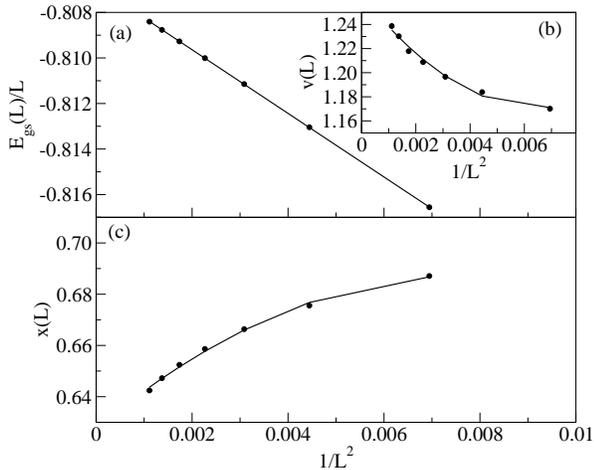}
  \caption{\label{fig:FSS}
  (a) Finite size scaling of the ground state energy $E_g(L)$.
  (b) Finite size scaling of the spin-wave velocity $v(L)$.
  (c) Finite size scaling of the the scaling dimension $x(L)$,
  which is obtained from excited state energies by applying the sum rule as described in the text.
  }
\end{figure}


Before studying the excited states, we first use the finite-size scaling of the ground state energy to estimate the central charge $c$.
According to CFT, for a system with size $L$, the ground state energy $E_0(L)$ should scale as \cite{Cardy:1986,Ludwig:1987gs}
\begin{equation}
  \frac{E_0(L)}{L}=\epsilon_\infty - \frac{\pi}{6L^2}cv,
  \label{eq:scaling_d_gs}
\end{equation}
where $\epsilon_\infty$ is the ground state energy per site in the thermodynamics limit
 and $v$ is the spin-wave velocity.
In Fig.~\ref{fig:FSS}(a) we show the finite size scaling of the ground state energy $E_g(L)$ with $L=12 - 30$,
from which we find $\frac{\pi}{6}cv=1.3968$. To find $c$ we need to estimate the
value of $v$, which is determined by the energy of the state at $k=2\pi/L$.
In order to decide which state at $k=2\pi/L$ should be used, we resort to the excitation spectrum of the spin-1 BB model.
For the spin-1 BB model at the SU(3) point, we find that at $k=2\pi/L$ the $S=1,2$ states are degenerate,
indicating that the spin-wave excitation belongs to the $\mathbf{8}$ or $\mathbf{\bar{8}}$ representation of the SU(3) group \cite{sup}.
In contrast, for our spin-2 model the $S=1,2$ states at $k=2\pi/L$ are split due to the absence of SU(3) symmetry for the Hamiltonian.
In this work we use the $S=2$ state at $k=2\pi/L$ to define the spin-wave velocity
(Alternative choices only make minor differences, see \cite{sup})
and set the energy scale in Fig.~\ref{fig:ED_L15}.
It corresponds to the first excited state in the $S=2$ sector and its energy can be obtained accurately by non-Abelian DMRG.
In Fig.~\ref{fig:FSS}(b) we show the $L$-dependent velocity $v(L)$ as a function of $L$.
By extrapolation $v(L)$ as $v+a/L^2+b/L^4$ we find $v=1.2643$.
Combined with the value of $cv$ above we find that $c=2.11$
which is very close to the expected value of $c=2$ for the $\text{SU(3)}_1$ WZW model \cite{sup}.

To further support the CFT to be the SU(3)$_1$ WZW model we estimate the scaling dimensions $x_i$
which are related to the scaling of the excited state energies $E_i(L)$ as
\begin{equation}
  \frac{E_i(L)-E_0(L)}{L}=\frac{2\pi v}{L^2} \left( x_i + \frac{d_i}{\ln L} \right),
\end{equation}
where $x_i=h_L+h_R$ and $d_i$ is the coefficient of the logarithmic correction due to the marginal operator.
Here $h_L=h^0_L+m_L$ and $h_R=h^0_R+m_R$, where $h^0_L$, $h^0_R$
correspond to the holomorphic and antiholomorphic conformal weights of the primary fields,
and $m_L$ and $m_R$ are non-negative integers describing descendant fields.
When the system size is a multiple of three, the lowest energy states at $k = \pm 2 \pi/3$ are expected to
belong to the representation $\mathbf{3\times \bar{3}}$ with $h_L=h_R=1/3$ \cite{Aguado:2009dma}.
Since the Hamiltonian has $\textrm{SU(2)}$ symmetry but not the $\textrm{SU(3)}$ symmetry,
the excited state energies split according to their SU(2) spin $S$.
Furthermore, states with different $S$ will pick up a different logarithmic correction $d_S$.
Fortunately, they can be removed by using the sum rule $\sum_S (2S+1)d_S=0$ \cite{Itoi:1997tv}.
By using the appropriate average of the excited states,
one can define an $L$-dependent scaling dimension $x(L)$ \cite{sup}.
In Fig.~\ref{fig:FSS}(c) we show $x(L)$ as a function of $L$.
By extrapolating $x(L)$ as $x+a/L^2+b/L^4$ we find $x=0.628$.
This agrees well with the expected value $x=2/3$ of the SU(3)$_1$ WZW model.
The above results on the central charge $c$ and scaling dimension $x$ confirm the CFT to be the SU(3)$_1$ WZW model.
One can argue that another possibility is the SU(3)$_2$ model.
However it has $c=3.2$, which make it unlikely to be the correct CFT.
Furthermore, we find that the excitation spectrum of the SU(3)$_2$ model
is not compatible with our ED result. The SU(3)$_2$ model thus can be safely excluded.
In calculations above we keep $m=2,800$ states of the SU(2) reduced basis.
This is equivalent to about $25,000$ standard DMRG states.
We have checked that the $m$ is big enough to ensure that the results are
in the finite size scaling regime as distinct from the finite entanglement regime \cite{sup}.

\begin{figure}
\includegraphics[width=0.9\columnwidth]{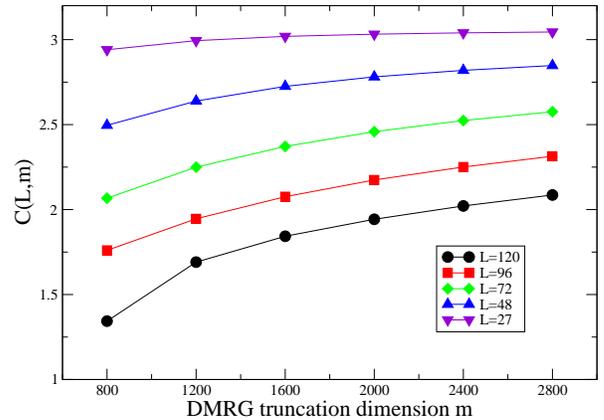}
\caption{\label{fig:DMRG}
  (Color online) $c(L,m)$ as a function of DMRG truncation number $m$ of SU(2) states for various systems sizes $L$
  with $\vec{\epsilon}=(0,-1,0)$. The $m=2,800$ states are equivalent to at least about 25,000 states of standard DMRG.
}
\end{figure}

In the above we obtained the central charge $c$ by considering the ground state energy.
In recent years the finite size scaling of the entanglement entropy (EE)
instead has been used intensively to  estimate the central charge.
Consider a system with periodic boundary conditions,
the EE of a subsystem of size $l$ is the von Neumann entropy
of the reduced density matrix $\rho_l$ of the subsystem; $S(l)=-\text{Tr}\left( \rho_l \log \rho_l \right)$.
It is known that for 1D conformal invariant critical system of size $L$,
the EE scales asymptotically as \cite{Calabrese:2004hl}
\begin{equation}
  S(l,L)=\frac{c}{3} \log\left[ \frac{L}{\pi} \sin\left(\frac{\pi l}{L} \right)  \right]+S_0.
  \label{eqn:s_scaling}
\end{equation}
where $c$ is the central charge of the CFT and $S_0$ is a non-universal constant.
Within DMRG it is straightforward to calculate $S(l,L)$ once the optimized ground state is obtained.
The accuracy of the DMRG calculation is controlled by the truncation dimension $m$ and the result
is numerically exact in the limit of $m\rightarrow \infty$.
For a given pair of $L$ and $m$ we fit Eq.~\ref{eqn:s_scaling} to obtain an effective central charge $c(L,m)$.
In Fig.~\ref{fig:DMRG} we plot $c(L,m)$ as a function of $m$ for several system sizes.
To our surprise we observe enormous finite size and truncation effects.
Similar phenomenon is also reported for critical $S=1/2$ XXZ chain, but only when
the system is extremely close to the ferromagnetic boundary with an emerging ferromagnetic length scale \cite{Chen:2013cy}.
Here the we have set the system to be far away from the ferromagnetic boundary but a pronounced effect is still observed.
In general $c(L,m)$ is a decreasing function of $L$ but an increasing function of $m$.
The true central charge is obtained in the limit of $c=c(L\rightarrow \infty, m\rightarrow \infty)$,
while $c(L,m\rightarrow \infty)$ provides a upper bound for the $c$.
Due to the enormous finite-size and truncation effects, we find it difficult to accurately determine
the value of $c$ from EE scaling, but certain bounds can be estimated.
From the smaller size data where $c(L,m)$ already saturates as m increases, we find strong evidence that $c \lessapprox 3$.
This again excludes the SU(3)$_2$ model with $c=3.2$.
For the largest $L=120$ used we find that $c(L, 2800) >2 $, which is consistent with the SU$_1$(3) model with $c=2$.
For even larger $L$ it is expected that even larger $m  \gg 2,800$  ($m \ge 25,000$ standard DMRG states) is needed, 
which is however beyond the typical size of DMRG calculations.

Some comments are now in order:
First, we find that there are similarities between the extended critical phase of the spin-2 model
and the extended critical period-three phase of the spin-1 BB model  \cite{Itoi:1997tv}.
For both models there are period-three structures and gapped trimerized states are formed for finite-size chain.
Furthermore, the corresponding critical theory is the SU(3)$_1$ WZW model in both cases.
There are, however, some crucial differences.
The spin-1 BB model has an enlarged SU(3) symmetry
when the strength of bi-linear and bi-quadractic terms are the same,
but the spin-2 model is never SU(3) symmetric within the phase space available,
except when $\epsilon_{0,2,4}$ are all equal where one has SU(5) symmetry.
It is only in the thermodynamic limit that the SU(3) symmetry emerges.
Furthermore, the critical phase of the spin-1 BB model is not accessible
from spin-1 bosons in lattice, while for spin-2 model the critical phase is accessible from spin-2 bosons.

Second, we observe that by using the finite-size scaling of the energies, data from smaller sizes are enough to precisely identify the CFT.
In contrast, it is difficult to identify the CFT via the finite-size scaling of the EE with typical computational resources.
The physical picture is as follows: Due to the proximity to the ferromagnetic phase,
there is a competition between the conformally invariant state and permutation symmetric state.
Two kind of states have very different EE scaling behavior, leading to enormous finite size and truncation effects  \cite{Chen:2013cy}.
Analysis based on energy scaling is less sensitive to such a competition
and accurate results for central charge and scaling dimensions can be obtained from smaller sizes data.
Our picture is also consistent with the existence of anomalous low energy states beyond the CFT prediction,
for example, the $S=3,4$ states at $k=0$ in Fig.~\ref{fig:ED_L15}.
They appear because at small length scale the system looks ferromagnetic.
While their energies are lower than the states associated with the primary field at small sizes,
it is expected that for larger sizes their re-scaled energy will move up to higher energies and become irrelevant.
In contrast  re-scaled energy of the states at $k=2\pi/3$ will converge to $2/3$ when SU(3) symmetry emerges
at larger sizes.

It is also natural to ask what kind of experimental signature can be observed.
Due to the non-local nature of the EE, it is not easy to measure the EE directly.
Recently there are proposal to measure the related quantities, the R\'enyi entropies,
within the cold atom framework \cite{Daley:2012bd}. It has also been shown that
the influence of the ferromagnetic state and the value of the effective central charge can
be detected and measured via R\'enyi entropies \cite{spin2_FM}.
It is thus in principle possible to experimentally verify the findings here with cold atom experiments.

In summary we study the critical phase of the spin-2 model obtained from spin-2 bosons in 1D lattice.
By using multiple approaches we identify the critical theory to be SU(3)$_1$ WZW model.
The Hamiltonian is never SU(3) symmetric but the SU(3) symmetry emerges in the thermodynamic limit.

We acknowledge inspiring conversations with M. A. Cazalilla. This research was supported by the NSC and NCTS of Taiwan. Ian McCulloch acknowledges financial support from the Australian Research Council Center for Engineered Quantum Systems.

\end{document}


\section{Supplementary materials}
In this supplementary material we provide additional information on the numerical results.
In particular we show results for the low energy spectra for the spin-1 bi-linear bi-quadratic (BB) model.
We also show some detail about the fitting of the central charge, spin-wave velocity, and scaling dimension
for the spin-2 model.
Additionally we show some numerical checks on the finite-size versus finite-entanglement scaling.
\subsection{Exact diagonalization for the  spin-1 model}

We first show the low energy excitation spectrum of the spin-1 BB model in the critical period-three phase.
While the model and the critical period-three phase are well studied and understood \cite{Fath:1993jk,Itoi:1997tv,Lauchli:2006ip},
here we provide the spectra as references to the spectra of the spin-2 model.
The Hamiltonian of the spin-1 BB model reads
\begin{equation}
  H=\sum_{i} \cos(\theta) \mathbf{S}_i\cdot \mathbf{S}_{i+1} + \sin(\theta)\left( \mathbf{S}_i\cdot \mathbf{S}_{i+1} \right)^2.
\end{equation}
The system is in the critical period-three phase when $\pi/4 \le \theta \le \pi/2$.
When $\theta=\pi/4$ the Hamiltonian becomes exactly solvable spin-1 Uimin-Lai-Sutherland (ULS) model,
with an enlarged SU(3) symmetry. It is known that the low energy physics of the whole critical period-three
regime is described by the $\textrm{SU(3)}_1$ Wess-Zumino-Witten (WZW) model.

\begin{figure}
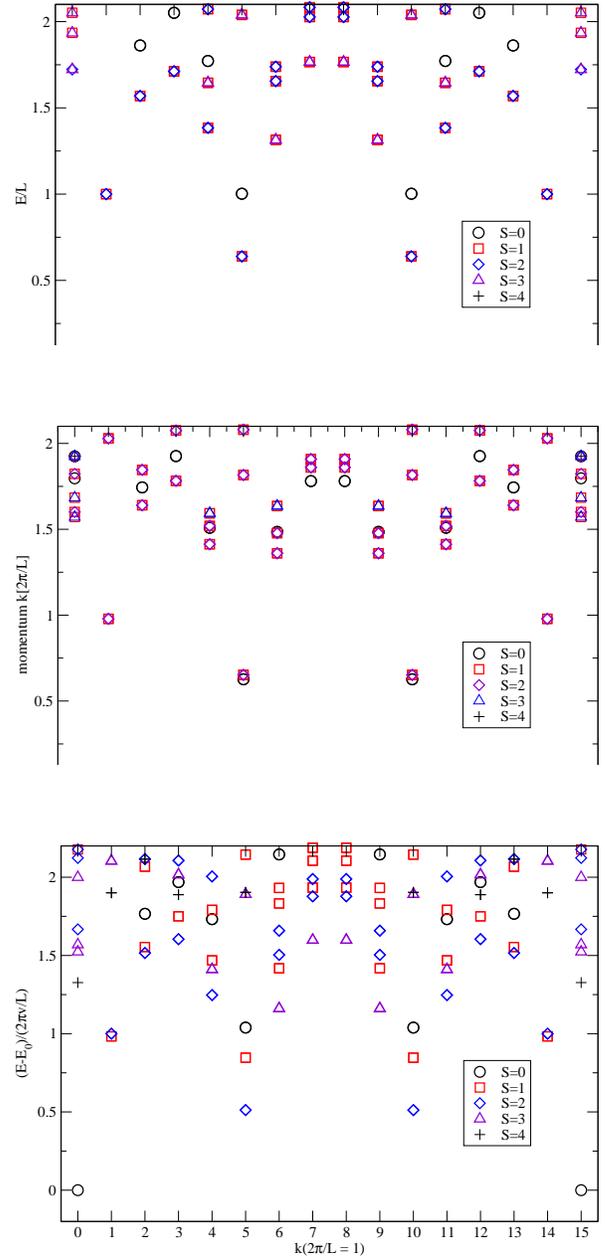

  \includegraphics[width=0.9\columnwidth]{ED1_0_SU3k1_L15.eps}
  \includegraphics[width=0.9\columnwidth]{ED1_0565_SU3k1_L15.eps}
  \includegraphics[width=0.9\columnwidth]{ED1_0_045_L15.eps}
  \caption{\label{fig:ED}
  Rescaled energy spectrum for spin-1 BB model at
  (a) $\theta=\pi/4$.
  (b) $\theta=\pi/4$ with $H^\prime=H+0.656H_2$.
  (c) $\theta=0.45\pi$.
  All data are obtained by  exact diagonalization for L=15 with periodic boundary conditions.
  }
\end{figure}

In Fig.\ref{fig:ED}(a) we show the rescaled energy spectrum for $\theta=\pi/4$
with system size $L=15$ using exact diagonalization. A period-three structure is clearly observed.
At $k=2\pi/3,4\pi/3$ there are soft modes with $S=0,1,2$. Furthermore the $S=1$ and $S=2$ states
are degenerate. For $L=3M$, where $M$ is an integer, it is expected that the holomorphic (antiholomorphic) part of
the  primary field should transform according to the $\mathbf{3}$ ($\mathbf{\bar{3}}$) representation of the SU(3) group.
While this implies that the states at these momenta would be $9$ fold
degenerate at $L \to \infty$, the presence of marginal operators split these states for finite $L$.
Since $\mathbf{3} \otimes \mathbf{\bar{3}}=\mathbf{1}\oplus\mathbf{8}$,
the $\mathbf{1}$ becomes the $S=0$ state while the $\mathbf{8}$ will be composed of $S=1$ and $S=2$ states
which are still degenerate if $\theta = \pi/4$.
Moreover, the $S=1$ and $S=2$ states will further split when $\theta$ is moved away from $\pi/4$
as the system loses the SU(3) symmetry. This is clearly observed in Fig.\ref{fig:ED}(c), where $\theta=0.45\pi$ is used.

The splitting due the marginal operator can be canceled by adding other perturbation Hamiltonian.
In particular one can add next nearest neighbor interaction $H_2$
\begin{equation}
  H_2=\sum_{i} \cos(\theta) \mathbf{S}_i\cdot \mathbf{S}_{i+2} + \sin(\theta)\left( \mathbf{S}_i\cdot \mathbf{S}_{i+2} \right)^2
\end{equation}
with appropriate strength to cancel the splitting at $\theta = \pi/4$.
In Fig.\ref{fig:ED}(b) we show the rescaled energy spectrum with $H^\prime=H+0.656H_2$.
We observe that all the state at $k=2\pi/3$ are nearly degenerate with 9 states in total.
Addition of this next-nearest neighbor interaction is suggested
by the fact that this term brings our spin chain
to the phase boundary point separating the gapless phase and
the gapped trimerized phase \cite{Corboz:2007cf}.  At this point,
the interaction parameter responsible for generating the logarithmic
corrections vanishes  (see Fig 1 in \cite{Itoi:1997tv}),
in analogy to the situation in spin-1/2 Heisenberg spin-chain \cite{Kitazawa:1996ea}.

Another key observation is that the lowest energy states at the momentum $k=2\pi/L$
are degenerate states with $S=1$ and $S=2$.
This indicates that the elementary excitation should have SU(3) representation $\mathbf{8}$.
The energy difference between this state and the ground state set the energy scale and the spin-wave velocity.
When the system is away from the SU(3) symmetry point, however, the  $S=1$ and $S=2$ states at $k=2\pi/L$ will split
as shown in Fig.\ref{fig:ED}(c) where $\theta=0.45\pi$ is used.
In all the plot we have use the energy difference between the $S=2$ state at $k=2\pi/L$ and the ground state
as the energy scale for the energy spectrum.

\subsection{Fitting of the spin-wave velocity, central charge, and scaling dimension for the spin-2 model}

In the following we provide the detail of the fitting procedure for the spin-wave velocity, central charge, and scaling dimension
for the spin-2 model. Data from $L=12$ to $L=30$ are used for the fitting. To be compatible with the period-three structure
we always set system size to be a multiple of three.

{\it Fitting for spin-wave velocity}:
To estimate the spin-wave velocity, we first calculate the ground state energy in the $S=0$ sector $E_0(S=0)$
and the first excited state energy in the $S=2$ sector $E_1(S=2)$. From the small size energy spectrum
of the spin-2 model and the spin-1 BB model shown in the main and supplementary text respectively we see
that this energy difference set the energy scale. At each $L$, we define a $L$-dependent velocity $v(L)$:
\begin{equation}
  v(L)=\frac{L}{2\pi} \left[ E_1(S=2)-E_0(S=0) \right].
\end{equation}
We then extrapolate $v(L)$ as \cite{2002PhRvB..65j4413H}
\begin{equation}
  v(L)=v+\frac{a}{L^2}+\frac{b}{L^4},
\end{equation}
as shown in Fig.~3(b) of the main text to obtain $v =1.2643$.
Alternatively, one can use the weighted average of the first excited state energies in the $S=1$ and $S=2$ sector
to estimate the spin-wave velocity. We have checked that this results in about $15\%$ difference
and it will not change any conclusion in the main text.

{\it Fitting for central charge}:
Since for ground state one has $h_L=h_R=0$, the ground state energy should scale as
\begin{equation}
  \frac{E_0(S=0,L)}{L}=\epsilon_\infty - \frac{\pi}{6L^2}cv,
  \label{eq:scaling_gs}
\end{equation}
where $\epsilon_\infty$ is the energy per site in the thermodynamic limit.
By fitting $E_0(S=0,L)/L$ as a function of $1/L^2$ as show in Fig.~3(a) of the main text,
we find $\frac{\pi}{6}cv=1.3968$, from which we find $c=2.11$.

{\it Fitting for scaling dimension}:
According to the conformal field theory, the energy of the primary scales as
\begin{equation}
  E(L)=\epsilon_\infty L + \frac{2\pi v}{L}\left( -\frac{c}{12} + x \right),
  \label{eq:scaling_ex}
\end{equation}
where $x=h_L +h_R$ is the scaling dimension.
For $L=3M$ the primary field with smallest scaling dimension should have $h_L=h_R=1/3$.
From the small size energy spectrum, we find that they correspond to the lowest energy states in the $S=1,2$ sectors
and the first excited state in the $S=0$ sector. Their energy difference with respect to the ground state hence scales as
\begin{equation}
  \Delta E(S,L) = \frac{2\pi v}{L} \left( x  + \frac{d_S}{\ln L} \right),
\end{equation}
where $x=2/3$.
Here we also include the logarithmic correction that depends on the SU(2) spin $S$.
The leading logarithmic correction can be removed by using the sum rule $\sum_S (2S+1) d_S=0$ \cite{Itoi:1997tv}.
To proceed, we first define a weighted average
\begin{equation}
  \Delta E(L) \equiv \frac{1}{9} \sum_{S=0,1,2} (2S+1) \Delta E(S,L).
\end{equation}
To avoid using $v$ directly in the fitting procedure, we use the $L$-dependent velocity $v(L)$
to define a $L$-dependent scaling dimension:
\begin{equation}
  x(L) = \frac {L}{2\pi v(L)} \left( \Delta E(L) - E_0(S=0,L) \right).
\end{equation}
We then extrapolate $x(L)$ as \cite{2002PhRvB..65j4413H}
\begin{equation}
  x(L)=x+\frac{a}{L^2}+\frac{b}{L^4}
\end{equation}
as shown in Fig.~3(c) of the main text to obtain $x=0.628$.
The result is consistent with the expected value $x=2/3$.


\subsection{Finite-size versus finite-entanglement scaling}

Consider a DMRG calculation for a critical system with periodic boundary conditions. If $L$ is the length of the system and $m$
is the number of the reduced basis kept. Two possible scaling regimes exist:
The finite-size scaling (FSS) regime corresponds to taking $m\rightarrow \infty$ first
and then $L\rightarrow \infty$, whereas the finite-entanglement scaling (FES) regime
corresponds to taking $L\rightarrow \infty$ first then $m\rightarrow \infty$.
When the calculation is performed within the FSS regime, the finite size scaling equations,
Eq.\ref{eq:scaling_gs} and Eq.\ref{eq:scaling_ex}, are valid.
In contrast, when the system is in the FES regime, a different scaling relation hold:
\begin{equation}
  \frac{E_0(L,m)}{L} = \epsilon_\infty - \Delta \left(\frac{1}{m^\kappa}\right)^2,
  \label{eq:scaling_FES}
\end{equation}
where $\kappa$ is the exponent for the scaling of the effective correlation length and $\Delta$ is a non universal constant.

\begin{figure}
  \includegraphics[width=0.9\columnwidth]{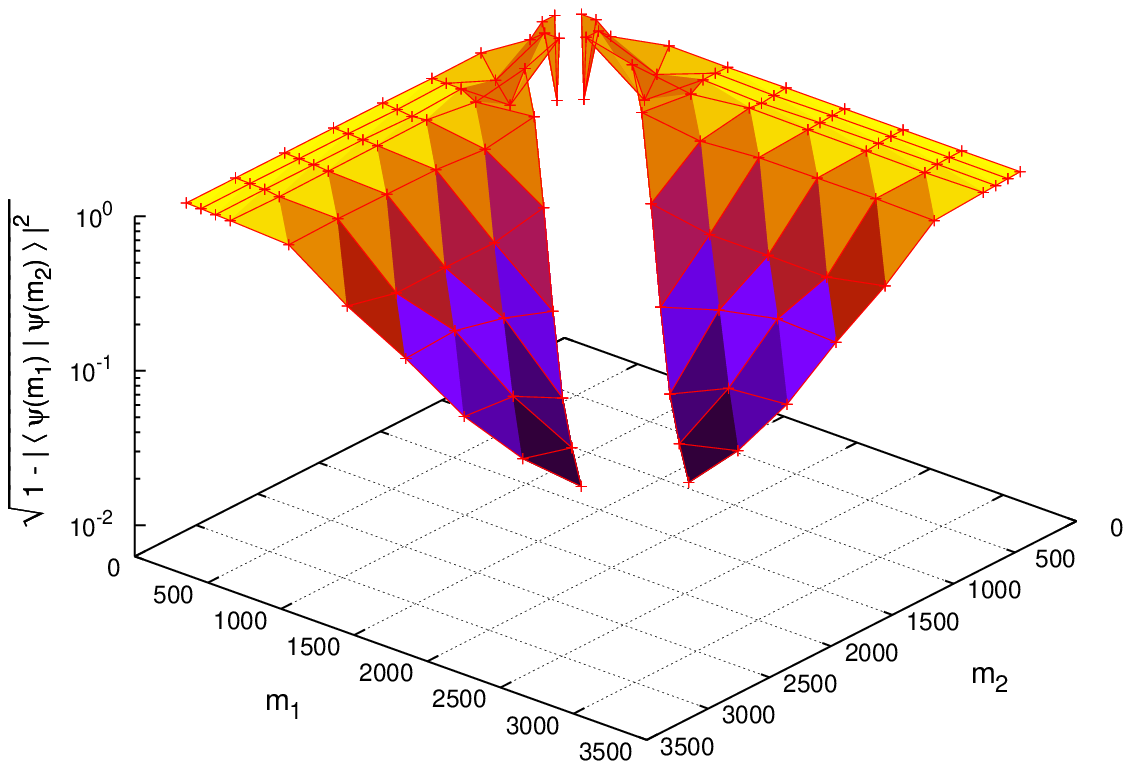}
  \caption{\label{fig:m1m2}
  (Color online) Trace distance between DMRG ground state with different truncation dimensions $m_1$ and $m_2$ for a $L=120$ chain.
  }
\end{figure}

In this work we rely on the FSS equations, Eq.~\ref{eq:scaling_gs} and Eq.~\ref{eq:scaling_ex},
to estimate the value of the central charge $c$ and the scaling dimension $x$.
It is thus important to ensure that the calculations are carried out in the predominately FSS regime.
In general the crossover between FSS and FES regimes in the parameter space $(L,m)$ is quite complex.
In Ref.~\cite{2012PhRvB..86g5117P} it is proposed that the minimal $m_r$ for faithful simulation in the FSS regime
can be estimated via monitoring the trace distance between ground-states obtained with different $m$'s.
It is expected that all DMRG ground states with $m \ge m_r$ have a much smaller trace distance
among each other than with DMRG ground states with $m < m_r$.
In Fig.~\ref{fig:m1m2} we plot the trace distance between DMRG ground state with different truncation dimensions
$m_1$ and $m_2$ for a $L=120$ chain. We observe that the for $m\le 800$ the trace distance among each other
is quite large, indicating that the calculation is in the FES regime. For $m \ge 2000$ we observe that the trance distance
among each other has become about two order of magnitude smaller, indicating that the calculation is in the FSS regime.
This ensures that the $m=2800$ used in this work is big enough to carry out the finite-size scaling analysis via  Eq.~\ref{eq:scaling_gs} and Eq.~\ref{eq:scaling_ex}.

%
%

